%% file: ppdpCyberSec.tex
\documentclass[conference]{IEEEtran}
\usepackage{times,epsfig, graphicx, subfigure, float}
\usepackage{url,endnotes}
\usepackage{algorithm}
\usepackage{algorithmic}
\usepackage{wrapfig,microtype}
\usepackage{times,setspace}
\usepackage{latexsym}
\usepackage{amsmath}
\usepackage{url}
\usepackage{mdwlist}
\usepackage{pifont}
\usepackage{fixltx2e}
\usepackage{flushend}
\usepackage{microtype,mdwlist}
\usepackage[font=small,labelfont=bf]{caption}

\begin{document}
%
\title{I Have the Proof: Providing Proofs of Past Data Possession in Cloud Forensics}

\author{\IEEEauthorblockN{Shams Zawoad}
\IEEEauthorblockA{University of Alabama at Birmingham\\
Birmingham, Alabama 35294-1170\\
Email: zawoad@cis.uab.edu}
\and
\IEEEauthorblockN{Ragib Hasan}
\IEEEauthorblockA{University of Alabama at Birmingham\\
Birmingham, Alabama 35294-1170\\
Email: ragib@cis.uab.edu}}

\maketitle

\begin{spacing}{0.95}
\input{abstract}
\IEEEpeerreviewmaketitle

\input{introduction}

\input{motivation}
\input{threatmodel}

\input{pds}
\input{discussion}
\input{relatedwork}
\input{conclusion}
\vspace{-11pt}
\input{ack}
\vspace{-11pt}
\end{spacing}

\begin{spacing}{0.95}
\begin{footnotesize} \bibliographystyle{IEEEtran}

\bibliography{../resource/cloudForensic_short}
\end{footnotesize}
\end{spacing}

\end{document}

%% file: abstract.tex
\begin{abstract}
Cloud computing has emerged as a popular computing paradigm in recent years. However, today's cloud computing architectures often lack support for computer forensic investigations. A key task of digital forensics is to prove the presence of a particular file in a given storage system. Unfortunately, it is very hard to do so in a cloud given the black-box nature of clouds and the multi-tenant cloud models. In clouds, analyzing the data from a virtual machine instance or data stored in a cloud storage only allows us to investigate the current content of the cloud storage, but not the previous contents. In this paper, we introduce the idea of building proofs of past data possession in the context of a cloud storage service. We present a scheme for creating such proofs and evaluate its performance in a real cloud provider. We also discuss how this proof of past data possession can be used effectively in cloud forensics.
\end{abstract}

%% file: introduction.tex
\section{Introduction}
\label{sec:introduction}
In the recent years, cloud computing has gained popularity as a high-performance and low-cost computing paradigm, which provides better utilization of resources using virtualization. Small and medium scale companies find cloud computing highly cost effective as it replaces the need of building costly physical and administrative infrastructure, and offers the pay-as-you-go structure for payment. These attractive features of cloud computing will increase both the private and federal cloud computing market intensively in near future.\cite{input2009,websitemarketresearchmedia}. 

However, the characteristics of cloud computing not only attract the business organizations, but also attract the malicious individuals to use clouds to evade the law. An attacker can target an application deployed in cloud, or she can use the cloud  to launch an attack. This new attack surface has opened a new area in digital forensics -- \emph{Cloud Forensics}. According to an annual report of Federal Bureau of Investigation (FBI), the size of the average digital forensic case is growing at 35\% per year in the United States. From 2003 to 2007, it increased from 83GB to 277 GB in 2007 \cite{fbi2008}. This rapid increase in digital forensics evidence drove the forensic experts to devise new techniques for digital forensics. At present, there are several established, proven digital forensics tools in the market. Unfortunately, because of the characteristics of cloud computing, many of the assumptions of digital forensics are not valid in cloud paradigm, e.g., in cloud environment we do not have the physical access to the evidence, which we take for granted in traditional privately owned computing system. Hence, cloud forensics brings new challenges from both technical and legal point of view and opens new research area for security and forensics experts.
	
The process of digital forensics starts with acquiring the digital evidence. In a cloud, the evidence could be the image of virtual machine, files stored in cloud storage, and logs provided by cloud service providers (CSP). But if an adversary shuts down the virtual machine or removes files from cloud storage, then there is no way to prove what file she possessed previously. To overcome this problem, Birk et al. propose continuous synchronization \cite{birk2011technicalIssues}. However, there is no solution which states how to do the continuous synchronization and how to prove the past data possession. 

In this paper, we take the first step towards building proofs of past data possession (PPDP). Such a proof can be used by forensic investigators to prove that a suspect stored a given file with a service provider at a past time period. To illustrate the problem, we present the following a hypothetical scenario:

\emph{Alice got access to some illegal insider trading \cite{insidertrading} information of XYZ corporation. Alice used this non-public information to gain illegal advantage in trading of stock and bonds. She hosted the file in a cloud storage and removed the file after trading to hide her tracks. After repeating this incident for several times, she got the attention of law enforcement agency and they assigned Bob to investigate the case. Bob acquired the possible insider trading information from XYZ corporation but could not prove whether Alice had access to and stored those files, as she deleted those after trading. If Alice used her personal computer to store the information, then the task for Bob would be easy. Using some current forensic tools, he could retrieve the deleted files or prove their existence in Alice's computer. But as Alice used the cloud storage, there was no way for Bob to prove the data possession. Cloud provider could store all the deleted files, but this would increase the storage cost extensively. Moreover, if cloud provider published the proof of the files insecurely, an intruder, Charley could get information of the files possessed by Alice. On the other hand, Alice might claim that Bob colluded with the cloud provider to prove a false allegation.} 	

To mitigate the challenges discussed in the above scenario, we propose the notion of a proof of \textit{past data possession} in this paper. 

\noindent\textbf{Contributions:~} The contributions of this paper are as follows:

\noindent\textbf{1)} We introduce the notion of a Proof of Past Data Possession (PPDP) in the context of digital forensics;

\noindent\textbf{2)} We propose an efficient and secured cryptographic scheme for creating a PPDP; and

\noindent\textbf{3)} We evaluate the proposed PPDP scheme using a commercial cloud vendor.

\noindent\textbf{Organization:~} The rest of this paper is organized as follows. Section~\ref{sec:motivation} provides some background informations and challenges of cloud forensics and section \ref{sec:threatmodel} describes the threat model. Section~\ref{sec:pds} presents our PPDP scheme, security analysis, and performance evaluation of the scheme in a real cloud environment. Section~\ref{sec:discussion} discusses the applicability of such proofs. Section~\ref{sec:relatedwork}, presents an overview of related research in cloud forensics, and finally, we conclude in Section~\ref{sec:conclusion}.

%% file: motivation.tex
\section{Background and Challenges}
\label{sec:motivation}
In this section, we present the definitions of digital forensics and cloud forensics, motivation of our work, and discuss about the challenges of cloud forensics.

\subsection{Digital Forensics}
Digital forensics is the process of preserving, collecting, confirming, identifying, analyzing, recording, and presenting crime scene information. Wolfe defines digital forensics as \emph{``A methodical series of techniques and procedures for gathering evidence, from computing equipment and various storage devices and digital media, that can be presented in a court of law in a coherent and meaningful format"} \cite{wiles2007best}.  According to a definition of NIST \cite{kent2006guide}, computer forensic is an applied science to identify a incident, collection, examination, and analysis of evidence data. While doing so, maintaining the integrity of the information and strict chain of custody for the data is mandatory. 

\subsection{Cloud forensics}
Cloud forensics can be defined as applying computer forensics procedures in cloud computing environment. As cloud computing is based on extensive network access, and as network forensics handles forensic investigation in private and public network, Ruan et al. defined cloud forensics as a subset of network forensic \cite{ruan2011cloud}. They also identified three dimensions in cloud forensics -- technical, organizational, and legal. Cloud forensics procedures will vary according to the service and deployment model of cloud computing. For Software-as-a-Service (SaaS) and Platform-as-a-Service (PaaS), we have very limited control over process or network monitoring. Whereas, we can gain more control in Infrastructure-as-a-Service (IaaS) and can deploy some forensic friendly logging mechanism. The first three steps of computer forensics will vary for different service and deployment model. For example, the evidence collection procedure of SaaS and IaaS will not be same.  For SaaS, we solely depend on the CSP to get the application log, while in IaaS, we can acquire the Virtual machine instance from the customer  and can enter into examination and analysis phase. On the other hand, in private deployment model, we have physical access to the digital evidence, but we merely can get physical access to public deployment model. 

\subsection{Motivation}
Though cloud computing offers numerous opportunities to different level of consumers, many security issues of cloud environment have not been resolved yet. According to a recent IDCI survey, 74\% of IT executives and CIO's referred security as the main reason to prevent their migration to the cloud services model \cite{websiteclavister}. Some recent and well-publicized attacks on cloud computing platform justify the concern with security, e.g., a botnet attack on Amazon's cloud infrastructure was reported in 2009 \cite{websiteamazon2009}. Besides attacking cloud infrastructure, adversaries can use the cloud  to launch attack on other systems. For example, an adversary can rent hundreds of virtual machines to launch a Distributed Denial of Service (DDoS) attack. After a successful attack, she can erase all the traces of the attack by turning off the virtual machines. A criminal can also keep her secret files (e.g., child pornography, terrorist documents) in cloud storage and can destroy all her local evidence to remain clean. When law enforcement investigates such a suspect, the suspect can deny having the illegal content in her cloud storage. At present, there is no way to claim that an adversary owns certain data at a given time. Researchers are working to protect the cloud environment from different types of attack. But in case of an attack, we also need to investigate the case, i.e., we need to carry out a digital forensic investigation in the cloud. Besides protecting the cloud, we need to focus on this issue. Unfortunately, there have been little research on adapting digital forensics for use in cloud environments. In this paper, we address this problem, which has significant real-life implications in law enforcement investigating cybercrime and terrorism.

\subsection{Challenges}
The inherent characteristics of a cloud have made it quite difficult to prove the possession of data at a past time period. For example, how can an investigator prove that the suspect stored an incriminating file last month in the cloud, a file she has recently deleted from the cloud storage? We must find secure techniques for creating such proofs of past data possession which will be admissible in a court of law as valid evidence. Many things can complicate such a proof. Clients may question the integrity of any such proofs, claiming that the forensic investigators or the prosecution and the CSP have colluded to plant an evidence in the cloud or have simply lied about the presence of the incriminating files in the cloud. The following reasons also make the PPDP challenging.

\noindent\textbf{Less Control in Clouds:~} In traditional computer forensics, the investigators have full control over the evidence (e.g., a hard drive confiscated by police). In a cloud, unfortunately, the control over data varies in different service models. Consumers have highest control in IaaS and least control in SaaS. This physical inaccessibility and lack of control over the system make evidence acquisition a challenging task in cloud forensics. For example, in our hypothetical case, Bob could have easily used traditional forensics tools to recover the deleted files if he had the physical access to cloud storage. Sometimes, it is even impossible to locate where the data actually reside physically. For this less control issue, we need to depend on the cloud service providers (CSP) for evidence acquisition. Which in turn brings the honesty issue of the CSP's employee, who is not a certified forensic investigator. Additionally, CSPs are not always obligated to provide all the necessary logs.


\noindent\textbf{Multi-tenancy:~} In cloud computing, multiple virtual machines (VM) can share the same physical infrastructure, i.e., data for multiple customers may be co-located. While generating the proof of data possession for one user, other users' data can be mingled with the proof. An alleged user can claim that the proof contains information of other user, not her. The investigator needs to prove it to the court that the proof indeed contains the information of the malicious user. Moreover, we need to preserve the privacy of the other tenants. 

\noindent\textbf{Chain of custody:~} Chain of custody should clearly depict how the evidence was collected, analyzed, and preserved in order to be presented as admissible evidence in court \cite{vacca2005computer}. In traditional forensic procedure, it starts with gaining the physical control of the evidence, e.g., computer, hard disk. However, in cloud forensics, this step is not possible due to the multi jurisdictional laws, procedures, and proprietary technology in cloud environment \cite{taylor2010digital,grisposcalm}. The PPDP must clarify certain things to maintain the chain of custody, e.g., how the proof was generated, stored, and accessed.

\noindent\textbf{Presentation:~} The final step of digital forensic investigation is presentation, where an investigator accumulates his findings and presents to the court as the evidence of a case. Challenges also lie in this step of cloud forensic \cite{reilly2011cloud}. If an investigator retrieves some deleted files from a personal computer, it will be easily presentable to the court. As we cannot retrieve the deleted files from the cloud storage, the proof of data possession must be presented to the court in a convenient way.

%% file: threatmodel.tex
\section{Threat Model}
\label{sec:threatmodel}
Before describing the threat model, we first define the important terms to clarify the threat model.
\vspace{-5pt}
\subsection{Definition of terms}
\begin{itemize}

\item \emph{User:} A user is a customer of the cloud service provider (CSP), who uses the CSP's storage service. A user can be malicious or honest.

\item \emph{Evidence:} An evidence can be a file, virtual machine image, log files, and any other digitally acquired data.

\item \emph{Proof of Past Data Possession (PPDP):} The PPDP contains the proof of data possession to verify whether the user actually possessed the evidence or not. 

\item \emph{CSP:} The Cloud Service Provider (CSP) will generate the PPDP and provide an web interface to users and investigators for verifying the evidence. 

\item \emph{Investigator:}  An investigator is a professional forensic investigator, who needs to verify the evidence from PPDP. An investigator can also be malicious or honest.

\item \emph{Intruder:} An intruder is a malicious person, who wants to reveal some information of the evidence from the PPDP.
\end{itemize}

\vspace{-5pt}
\subsection{Attacker's Capability}
In our threat model, we assume that the users and the investigators do not trust the CSP, and both of them can be malicious. A user can delete records from her storage but cannot change the PPDP by herself. An investigator can plant a false evidence only when he colludes with the CSP, and cannot change the PPDP by himself. A CSP can produce false PPDP only after colluding with a user or an investigator. The CSP can deny hosting any evidence or repudiate any published PPDP. An intruder can acquire the PPDPs of user to learn about the evidence or evidence change history from the proofs.

\vspace{-5pt}
\subsection{Possible Attacks}
There can be different types of attacks on past data possession. A user can deny any evidence ownership, an investigator can produce invalid proof, even a CSP can deny hosting a file. Below we mention the possible attacks:
\begin{itemize}
\item \emph{Denial of possession:} A user can delete file from her cloud storage. Later, if an investigator found other evidence of her deleted file, she denies to have the file. In a variation of this, a colluding investigator can also delete evidence from the user's storage and deny to find the evidence. 
\item \emph{False presence:} If an investigator is not trustworthy, he can plant a false evidence. A user can also present a false proof, which can make her free from the accusation.
\item \emph{Evidence contamination:} User and investigator can modify the evidence to prove their claim. 
\item \emph{Repudiation by CSP:} An otherwise honest CSP can deny hosting a file or can deny a published PPDP after-the-fact.
\item \emph{Repudiation by User:} As data are co-mingled in the cloud, a malicious user can claim that the published proof contains other cloud user's data.
\item \emph{Privacy violation:} As the CSP published the proof of past data possession publicly on the web, any malicious person can acquire the published PPDP and try to learn about the evidence from the proof. This attack can violates the privacy of user data hosted in the cloud. 
\end{itemize}

\subsection{System Property}
Our mechanism should prevent any malicious party to produce a false PPDP. A false PPDP attests the user's possession of record, which the user does not actually own. Once the proof has been published, the CSP can neither modify the proof nor repudiate any published proof. We must also prevent false implications by dishonest forensic investigators. Our system provides the following integrity and confidentiality properties:

\noindent\textbf{I1:} An investigator or user whether acting alone or colluding together cannot remove any evidence.

\noindent\textbf{I2:} Two colluding investigator and user, or investigator and user acting alone cannot plant any invalid evidence. 

\noindent\textbf{I3:} An investigator or user whether acting alone or colluding together cannot change any existing evidence.

\noindent\textbf{I4:} After publishing the proof of evidence, the CSP cannot deny hosting any evidence.

\noindent\textbf{I5:} The CSP cannot repudiate any previously published proof.

\noindent\textbf{C1:} From the published proof, no adversaries can recover any evidence.

\noindent\textbf{C2:} Adversaries will not be able to learn about the change history from the published proof.

%% file: pds.tex
\section{Proof of Past Data Possession}
\label{sec:pds}
In this section, we present the PPDP scheme, implementation, evaluation, and security analysis of the scheme.

\vspace{-5pt}
\subsection{The Scheme}
We propose our scheme based on Bloom filters. A Bloom filter is a probabilistic data structure with no false negatives rate, which is used to check whether an element is a member of a set or not \cite{bloom1970space}. Bloom filter stores the membership information in a bit array. Bloom filters decrease the element insertion time and membership checking time. The only drawback of the Bloom filter is the probability of finding false positives. However, we can decrease the false positive probability by using a large bit array. In our scheme, we maintain separate Bloom filters for each users.
\smallskip

\begin{figure}[!ht]
\centering
\includegraphics[width=0.48\textwidth]{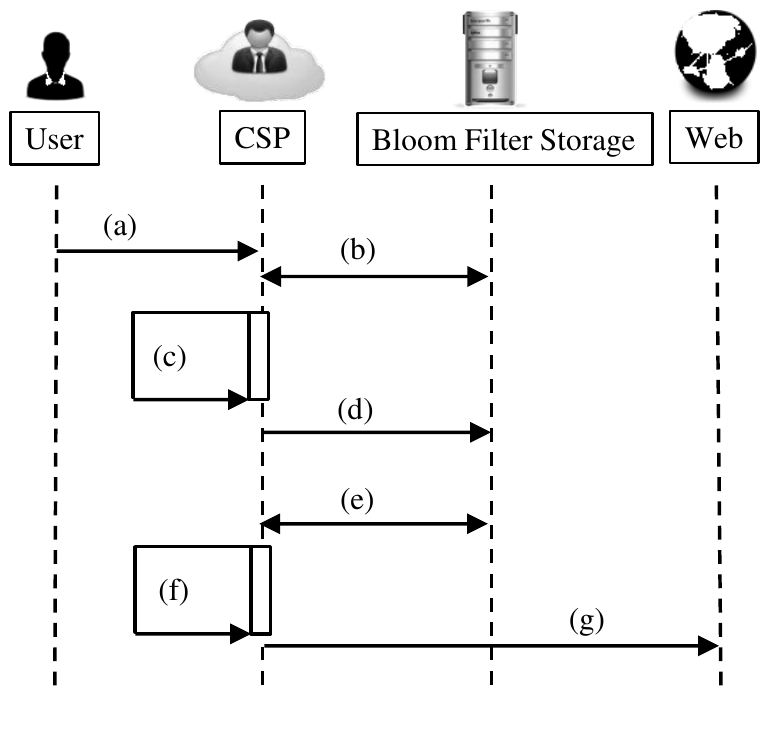}
\caption{PPDP Generation Process Flow}
\label{figure:ppdpflowshort}
\end{figure}

\noindent\textbf{Proof Insertion:~}
Figure \ref{figure:ppdpflowshort} shows the flow of PPDP generation and below is the description of the  process flow.

 \textbf{(a)} A user can upload a file to a cloud server, create a new file, or a new log file can be generated in cloud environment.

 \textbf{(b)} After acquiring any new or updated file, the CSP will get the last Bloom filter of that user, which is actually the bit array positions for all the previously inserted records of that user.

 \textbf{(c)} The CSP will then hash the file including the user's information, generate the bit positions from the hash value, and update the bloom filter with the latest bit positions.

 \textbf{(d)} Next, the CSP will store the updated bloom filter in the bloom filter storage.

 \textbf{(e)} At the end of each day, the CSP will retrieve the Bloom filter entry of each user. We denote this as \emph{DS\textsubscript{u}}.

 \textbf{(f)} Next, the CSP will hash the \emph{DS\textsubscript{u}} and sign it with its private key. This will be the \emph{PPDP\textsubscript{u}} of that day.
\begin{equation}
	PPDP\textsubscript{u} = \textless H(DS\textsubscript{u}), S\textsubscript{PKC}(H(DS\textsubscript{u})), t \textgreater
\end{equation}
Where \emph{t} is the proof generation time, \emph{H(DS\textsubscript{u})} represents the hash of \emph{DS\textsubscript{u}}, and S\textsubscript{PKC}(H(DS\textsubscript{u})) represents the signature on the \emph{H(DS\textsubscript{u})} using the private key of the CSP, \emph{PKC}.

 \textbf{(g)} After computing the proof of data possession, the CSP will publish the \emph{PPDP\textsubscript{u}} and its public key \emph{PK} on the web. These informations can also be available by RSS feed to protect it from manipulation by the CSP. The CSP will keep the \emph{DS\textsubscript{u}} and \emph{PKC} secret to itself.

A more advanced version of the above process can be used to prevent the repudiation by user attack for multi-tenancy case. As multiple users' data are co-located in same infrastructure, a user can claim that the proof contains information of her neighbour's file. To prevent this claim, all the evidence can be signed by the user's private key. Instead of hashing the file, the CSP now hashes the signature of the file to get the bit positions.  

The false positive probability is proportional to the number of elements contained in the Bloom filter. There will be a change in the flow when the total number of evidence crosses the expected number of elements for the Bloom filter. If the expected number of elements is \emph{n}, total number of bits required is \emph{m}, and number of hash functions is \emph{k} then the false positive probability \emph{p} is determined by the following equation \cite{bloomequation}.
\begin{equation}
	p = (1 - e\textsuperscript{$\frac{-k*n}{m}$})^k
\end{equation}

According to the above equation, if the number of elements increases, the false positive probability will also increase. Eventually, increasing false positive probability will increase the chance of planting false evidence attack. There are two options to mitigate this problem:

\noindent \emph{Option A:} After acquiring a file, the CSP will always check whether the current number of elements crosses the expected number of elements for the new file. If it does not cross the limit, then we can continue from step (b) of the above process flow. On the other hand, if it crosses the limit, then without getting the last bloom filter of the user, the CSP will create a new empty Bloom filter for the user and continue from step (c) using the newly created Bloom filter.

\noindent \emph{Option B:} As the above option add an extra checking before each proof insertion, it will increase the overhead of proof insertion. Another option can be creating a new Bloom filter of a user after a certain time. If a user creates \emph{w} number of files on average in every month then we can create a new Bloom filter for her after every \emph{$\frac{n}{w}$} months. Using a fixed amount of time for creating new empty Bloom filter can generate two problems -- for a user with low \emph{w}, the space will be misused and for a user with high \emph{w}, the false positive probability will increase. So creating a new Bloom filter after every \emph{$\frac{n}{w}$} months can balance the space requirement and the false positive probability.

\begin{figure}[!ht]
\centering
\includegraphics[width=0.48\textwidth]{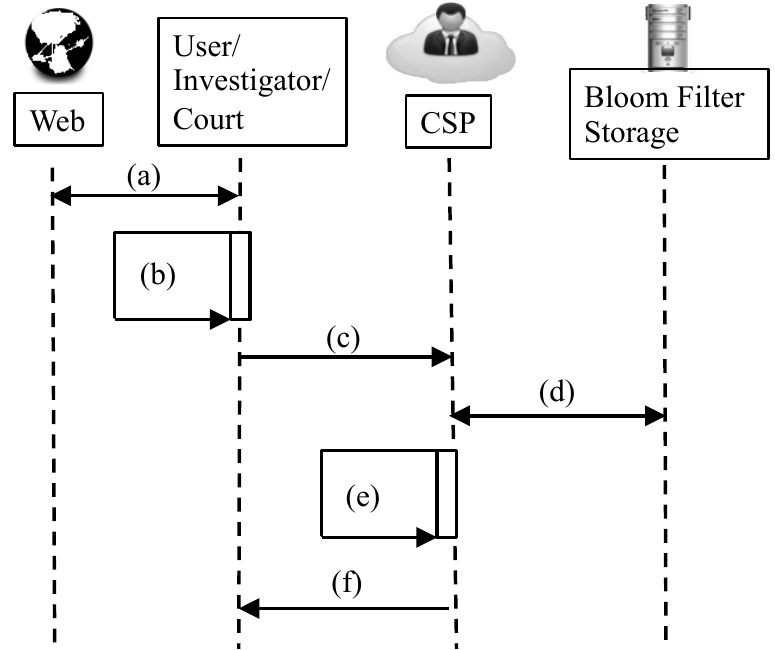}
\caption{Evidence Verification Process Flow}
\label{figure:ppdpverification}
\end{figure}

\smallskip
\noindent\textbf{Verification:~} Figure \ref{figure:ppdpverification} shows the verification process flow and below is the details of the process. 

 \textbf{(a)} User, investigator, or court -- any party can verify the validity of evidence. In the first step, they take the published \emph{PPDP\textsubscript{u}} from the web.

 \textbf{(b)} The verifying party will then check the validity of the \emph{PPDP\textsubscript{u}}. To check the validity, they first decrypt the signature of the hashed bloom filter by using the public key of the CSP. If the decrypted value and the hashed bloom filter are same, then they will proceed to next step.

 \textbf{(c)} Next, the verifying party will upload the file with the user information from the management panel of the CSP.

 \textbf{(d)} The CSP will then pick the last bloom filter entry, the hashed value of which was published on the web.

 \textbf{(e)} Next, the CSP will hash the uploaded evidence with the user information, calculate the bits positions, and compare these bit positions with the retrieved Bloom filter entry. If all the calculated bit positions are set in the latest Bloom filter entry, we can say that the user actually possessed the file. One single false bit position means the evidence is not valid.

 \textbf{(f)} After comparison, the CSP reports a positive or negative matching result to the verifying party. \smallskip

\textbf{Evidence Time-line Analysis:~} As the CSP publishes the PPDP everyday, we can identify the generation time of evidence. We can either identify the exact generation time or a time range in which an evidence was present.

To identify the generation time of an evidence, investigator will first send the evidence and user information to the CSP. The CSP will then acquire all the Bloom filter of the user and start searching to get a positive match. Positive match can be found for multiple PPDPs, but the generation time of the oldest PPDP will be the  generation time for the evidence.

Investigator or court may also want to know whether a evidence was present in a particular time range. In that case, they will send the evidence, user information, and the desired time range to the CSP. After acquiring all the necessary informations, the CSP will retrieve the Bloom filters of that user, which lie in the particular date range. The CSP then starts a binary search to find a positive match. If a positive match is found, then we can say that the evidence was present in the selected time range.

\subsection{Security Analysis}
In our collusion model, there are three entities involved -- CSP, user, and Investigator. All of them can be malicious individually or can collude with each other. We denote an honest CSP as C, a dishonest CSP as $\bar{C}$, an honest user as U, a dishonest user as $\bar{U}$, an honest investigator as I, and a dishonest investigator as $\bar{I}$. Hence, there can be total eight possible combination. 

\noindent\textbf{CUI} For the first case, where everybody is honest, there is nothing to worry about. In that case, there is no chance of evidence contamination. \smallskip

\noindent\textbf{$\bar{\textbf{C}}$UI} A CSP can produce incorrect PPDP, which is not a feasible scenario, as there is no benefit for CSP to be malicious alone. \smallskip

\noindent\textbf{C$\bar{\textbf{U}}$I} In this case, where only the user is dishonest, she can delete a file from his cloud storage and later claim that she did not have that file. For example, a malicious user might have hosted a child pornography site, and later she deleted all the evidence. If the investigator comes with some contraband images, then using our scheme he will be able to verify whether the user actually possessed those files or not. The CSP picks up the last Bloom filter entry, generates the hash of the evidence that the investigator provided, and by comparing it with the retrieved Bloom filter it can say whether the user is honest or not. It proves our claim for the I1, I2, and I3 integrity properties when the user is malicious alone.\smallskip

\noindent\textbf{CU$\bar{\textbf{I}}$} In this scenario, where only the investigator is dishonest, she can come up with a false evidence to frame an honest user. Our system will prevent this case from happening. Using the last Bloom filter of the user, the CSP can easily verify that the false evidence does not actually present in user's storage. A dishonest investigator can also deny a valid evidence, but using our system, the users will be able to establish their claim. Hence, we can ensure the I1, I2, and I3 integrity properties when the investigator is malicious alone.\smallskip

\noindent\textbf{C$\bar{\textbf{U}}\bar{\textbf{I}}$} In this case, a malicious user can collude with a dishonest investigator to make herself free from an accusation. However, as long as the CSP does not collude with them, user or investigator cannot prove their false claim. If user and investigator agree on removing an evidence, they cannot prove it to court as long as the PPDP is valid. Thus, our system ensures the I1, I2, and I3 integrity properties when the investigator and the user collude.\smallskip

\noindent\textbf{$\bar{\textbf{C}}$U$\bar{\textbf{I}}$} A dishonest CSP and a malicious investigator can collude to plant a false proof. In that case, the CSP was honest before colluding with the investigator, but becomes dishonest afterwards. Which means, the proof generated previously were correct but the proof generated after the colluding point will be incorrect. Here, we are proposing a scheme for proof of past data possession, that means the proof were valid at the time of generation so that nobody can establish an invalid evidence after a valid proof generation. Using the advance version of PPDP generation, we can even block the future false proof generation for this collusion combination. A CSP or an investigator cannot sign the file by user's private key. Therefore, storing the signature of the file in the Bloom filter can prevent producing any false evidence by a colluding CSP and an investigator.\smallskip

\noindent\textbf{$\bar{\textbf{C}}\bar{\textbf{U}}$I} A dishonest CSP and a malicious user can collude together to remove valid evidence or plant invalid evidence. Before colluding with the user, the CSP generated valid proof. This proof can detect any deletion of past data. However, after colluding with the user, CSP can produce invalid proof. This invalid proof will still detect the removal of data, which was in user's storage before the colluding point but will not work for the data, which the user owned after collusion.  \smallskip

\noindent\textbf{$\bar{\textbf{C}}\bar{\textbf{U}}\bar{\textbf{I}}$} Finally, a malicious user can collude with a CSP and an investigator to prove her honest. Even all of the three parties collude, they cannot come up with any invalid evidence for which a valid proof has already been published by the CSP. However, they can alter the new evidence when the CSP is dishonest and publishes invalid proof of the new evidence. \smallskip

In the last three cases, where the CSP is also dishonest we can ensure the I4 and I5 integrity properties by PPDP. As the published hash of bloom filter is signed with the CSP's private key, it cannot repudiate the proof (I5). After decrypting the signed value, if it matches with the hashed bloom filter value, the CSP cannot repudiate the published value. Additionally, if the CSP comes up with a false \emph{PPDP\textsubscript{u}} in place of a published PPDP\textsubscript{u}, then it will be easily detected. In that case, the \emph{H(DS\textsubscript{u})} of the published \emph{PPDP\textsubscript{u}} and the \emph{H(DS\textsubscript{u})} of the false \emph{PPDP\textsubscript{u}} will not be same. 

The published bloom filter contains the hash of all the previous evidence. If the published bloom filter contains the hash of a file, then the CSP cannot deny hosting that file (I4) when the probability of false positive rate is zero.

Besides the integrity properties, our system also ensures the two confidentiality properties -- C1 and C2, which will ensure cloud user's privacy. A CSP publishes the hashed value of the bloom filter and the signature of that. As the hash function ensures the one-way property, adversaries will not be able to know about the bloom filter content from the hash value (C1). As the CSP publishes the PPDP everyday, adversaries can try to know about the evidence change-history from the regularly published proof. However, it is not possible to know about the change-history from the hashed value.

We can defend the repudiation by user attack for multi-tenancy case by using the advanced version of PPDP generation, where the user's signature of the file is stored. As the key for the signature is private to the user, nobody other than the user can sign a file using the secret key. So, if a positive match found for a signature of an evidence, the user cannot repudiate the file possession.

\subsection{Evaluation}
To evaluate the feasibility and the performance of our scheme, we first set up a ftp server in an Amazon EC2 micro instance. Then we implemented our proof-of-concept application using JDK 1.6 and MySQL Community Server - 5.1.53. We upload files from a personal computer with Intel Core-i5-24305 CPU @ 2.40 GHz processor and 8GB RAM. We configured the Bloom filter with 0.01\% false positive probability for 1000 element and used seven hash functions to maintain this configuration. We used RSA (1024 bit) for signature generation and  SHA-1(160 bit) hash function for hashing. As the unique information of a user, we used the user's email address. We created a directory watcher, which watches the directories of each users. Whenever any new file is uploaded to user's directory, we hash that file along with the user information and insert into the latest bloom filter entry of that user.\smallskip

\begin{figure}[!ht]
\centering
\includegraphics[width=0.49\textwidth]{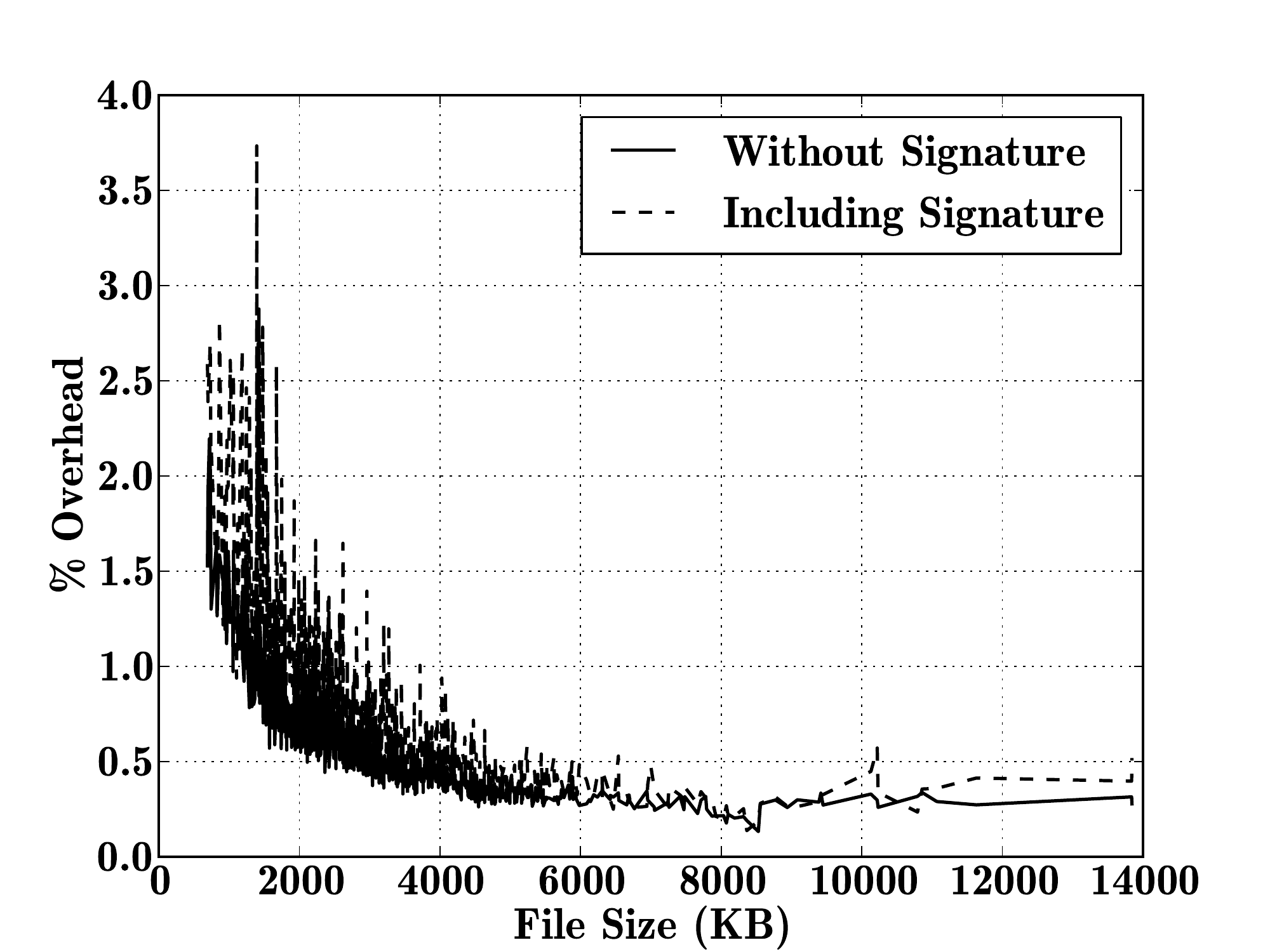}
\caption{\% Overhead associated with time needed to insert the PPDP. }
\label{figure:overhead}
\end{figure}

\noindent\textbf{Associated Overhead:} For PPDP generation, we follow the option B of the process flow, i.e., we did not check the the number of elements in bloom filter before inserting the proof. To check the overhead of inserting the proof of a file in our system, we first upload 1000 files to our ftp server and measure time for each file. File size was distributed in log normal distribution. The lowest and the highest file size was 693 KB and 13843 KB respectively. Then we again upload the files to the server, but now in a directory which is being watched by our directory watcher. This time, we calculate the time of each file including the time to insert the proof of the file. From the difference of the this two time, we calculate the \% overhead associated with inserting the proof of the file. We run our experiment for several times to measure the average overhead. Finally, we run the same experiment for signature-based Bloom filter. Figure \ref{figure:overhead} depicts our findings. Without signature, the overhead varies from 2.1959755\% to 0.132992189\% and when the file size crosses 4.5 MB, it varies from 0.2\% to 0.3\%, which is significantly low. For signature-based Bloom filter the overhead varies from 0.137635076\% to  3.733947515\% and after 6 MB, it varies between 0.3\% to 0.6\%. So there is a slight increase in overhead for the signature-based Bloom filter.\smallskip

\begin{figure}[!ht]
\centering
\includegraphics[width=0.45\textwidth]{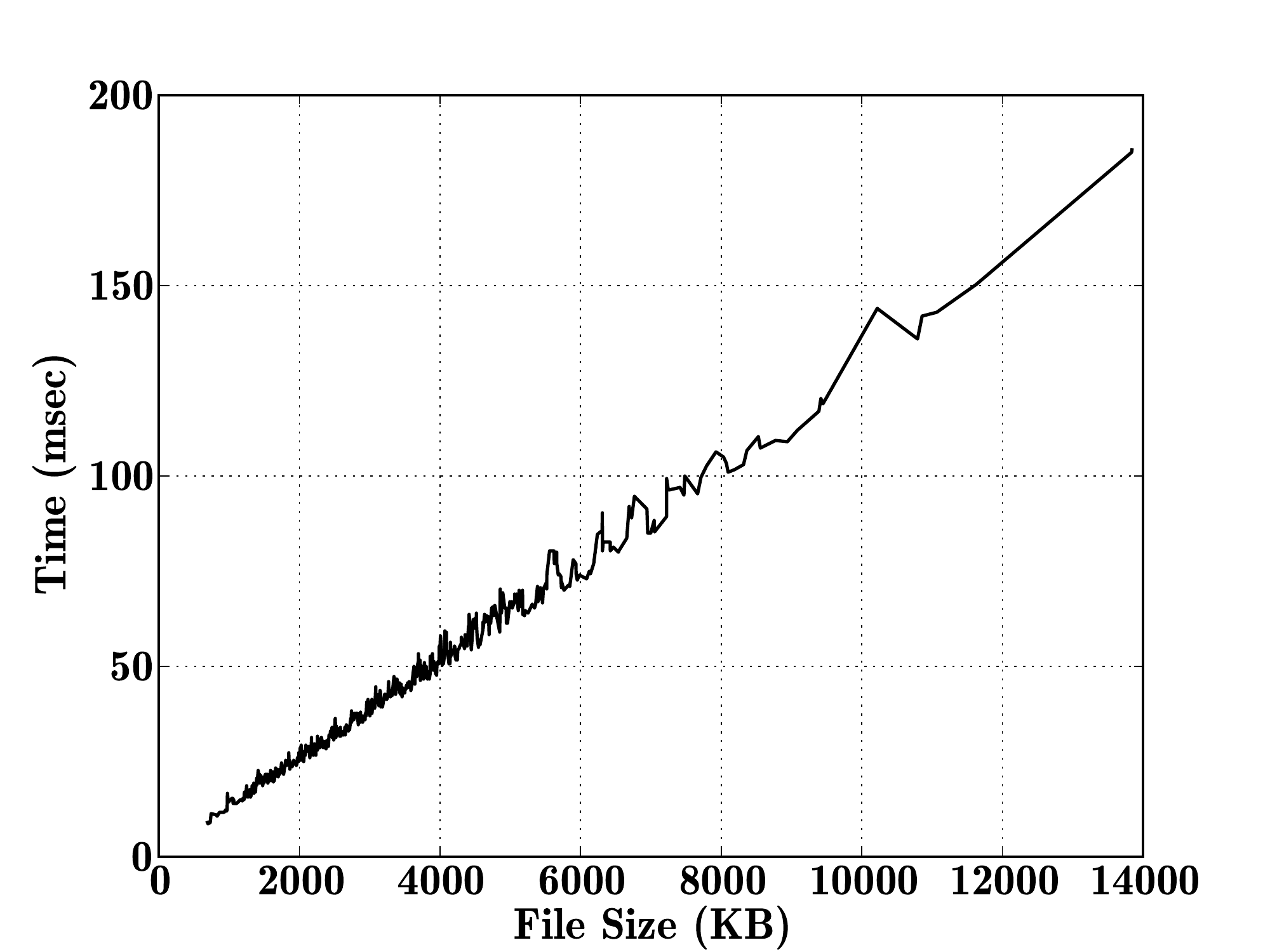}
\caption{Average time required for true negative matching.}
\label{figure:truenegative}
\end{figure}

\noindent\textbf{Proof Checking:} To measure the performance of proof checking, we consider both the true positive matching and true negative matching. By true negative matching, we mean that the file was not present in the PPDP and our system detects it correctly. And by true positive matching, we mean that the file was actually present in the PPDP and the application detects the presence correctly. To test the true negative matching time, we use the same file set, that we have used to measure the overhead. First, we choose such a PPDP, that the files are not present in that proof. Then we run the proof checking program with all the files and measure the time to find a true negative match. We run this test several times to get the average time. Figure \ref{figure:truenegative} depicts our result. From the graph, we can observe that the time increases with the file size. The more the file size is, the more time is required.

\begin{figure}[!ht]
\centering
\includegraphics[width=0.45\textwidth]{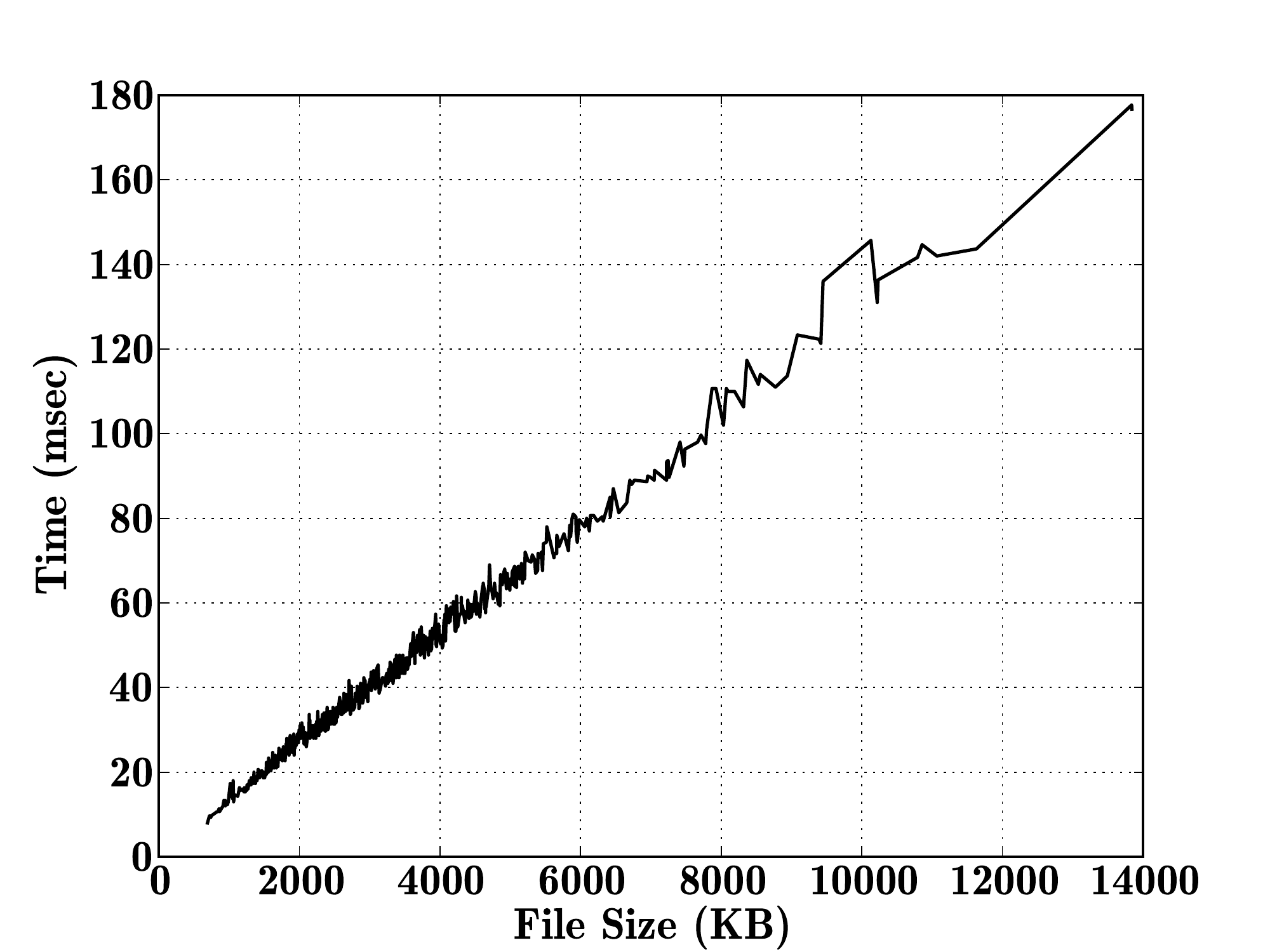}
\caption{Average time required for true positive matching.}
\label{figure:truepositive}
\end{figure}

Before measuring the true positive matching time, we first insert all the files in our system. That means the latest PPDP contains the proof of all the files. We select this last proof and run the proof checking program again for all the files. We find a true positive match for all of the files and note the time. Figure \ref{figure:truepositive} illustrates the average time for finding a true positive match for different file size. From the graph, we can state that the finding true positive matching time also depends on the file size. For larger file size we need higher time.

In both true positive and negative matching, the required time increases with the increasing file size. And the reason behind that is the hashing time. The time of hashing is proportional with the file size. In our program, to calculate the bit positions, we used 7 hash functions. Hence, it is obvious that the matching time will increase with the increase in file size.\smallskip

\noindent\textbf{Storage Overhead:} In our experiment, we used 0.01 false positive probability for 1000 elements, which require 10099 bits, i.e., 1262.375 byte. We maintain one bloom filter for one user. Also we are publishing the Bloom filter every day, i.e., we need 1262.375 byte storage for each user everyday. For \emph{n} number of users we need \emph{n*1262.375} byte storage everyday. This is significantly low, because implementing our scheme will ensure the preservation of data possession proof without preserving the file itself. For example, if on average every user deletes 5 mega bytes file daily, then everyday our scheme will save \emph{n*4.99} mega byte storage, but still preserves the proof of data possession.

%% file: discussion.tex
\section{Discussion}
\label{sec:discussion}
In this section, we discuss the applicability of PPDP scheme in cloud forensics and how it can contribute in build a regulatory compliant cloud.
\subsection{Application in Cloud Forensics}
The above scheme will be applicable in cloud forensics and help the investigators to justify the evidence. In our scheme, CSPs will play a vital role. When storing a file in disk, they need to do some additional work and need to provide a new feature in management console to publish and verify the proof of past data. Storing the proof will need some extra storage for the CSP, but on the other side they can earn money from forensic agency by providing forensics-as-a-service. As in this way, the CSPs do not need to store the evidence itself, only need to store the proof of the evidence, it will actually quite cost effective way to provide forensic-as-a-service. This is a continuous synchronization approach, but without preserving the data. If a malicious user deletes her secret data from the cloud environment, the CSPs can still preserve the proof of the record without preserving the data itself. Later, if an inspector gets the same data from any other source, he can check whether the data was owned by a particular malicious user or not. If a malicious investigator produces some false evidence, the user will also get the chance to prove it false. For cloud storage service, users will get the facility to verify the integrity of their data. If a user is suspicious about the integrity of a file, he can check the validity of the file from the management console. 

Publishing the signed hash of a Bloom filter also ensures the trustworthiness of the CSP. If any user or investigator objects about the honesty of the CSP, it can defence this claim by providing the Bloom filter content to the court. If the hash of this Bloom filter content is same as the published hash (\emph{H(DS)}), then we can treat the CSP as honest. If the CSP is not trustworthy and change the Bloom filter content, then the Bloom filter content submitted to the court cannot produce the same \emph{H(DS)}, which was signed and published previously.
\subsection{Regulatory Compliant Cloud}
The Sarbanes-Oxley (SOX) Act 2002 mandates public companies to provide disclosure and accountability of their financial reporting, subject to independent audits \cite{websitesoxact2002}. While this act introduced major changes to the regulation of financial practice and corporate governance, it also brings new challenges for cloud computing. The SOX act mandates that the financial information must be reside in an auditable storage, which the CSPs do not provide. Business organization cannot move their financial information to cloud infrastructure as it does not comply with the act. The SOX act mandates accurate financial disclosure \cite{websitecorporatefraudact}. However, in today's cloud infrastructure, auditors cannot verify the authenticity of financial reports provided by the corporations. PPDP-enabled cloud infrastructure can provide this functionality. Implementing our scheme will help the auditors to verify whether a corporation actually possesses the records or it produces false reports. Hence, implementing the PPDP scheme will be a big step towards building a SOX-compliant cloud. Building a SOX-compliant cloud will open a new business model in cloud computing and also the business organizations can reduce their investment in buying SOX-compliant storage.

%% file: relatedwork.tex
\section{Related Work}
\label{sec:relatedwork}
Cloud forensics is a relatively new topic. Several researchers have proposed solutions to overcome some of the challenges of cloud forensics. Delport et al. focused on isolating an instance to mitigate the multi-tenancy issue \cite{Delport2011cloud}. Isolation is necessary because it helps to protect evidence from contamination. The first proposed technique is instance relocation. The second technique is server farming, which can be used to re-routing the request between user and node. The third technique is failover, where there is at least one server that is replicating another. The last approach of isolating an instance is to place it in a sandbox. To overcome the problem of volatile data, Birk et al. mentioned about the possibility of continuous synchronization of the volatile data with a persistent storage \cite{birk2011technicalIssues}. They also proposed that the CSP can provide network, process, and access logs through a read-only API to get necessary logs from all the three cloud service models. However, they did not provide any guideline or any practical implementation of the procedures. 

Provenance provides the history of an object. By implementing secure provenance, we can get some important forensic information, e.g., who owns the data at a given time, who accesses the data and when. Secure provenance can be useful for ensuring the chain of custody in cloud forensics as it can provide the chronological access history of evidence, how it was analysed, and preserved. There have been some works for secure provenance in cloud computing \cite{muniswamy2010cloud,muniswamy2010provenance}, but no CSP has implemented this mechanism yet.


There are other important challenges orthogonal to our work. To overcome the data acquisition problem, Dykstra et al. recommended a cloud management plane for using in the IaaS model \cite{dykstraacquiring}. From the console panel, customers, as well as investigators can collect VM image, network, process, database logs, and other digital evidence, which cannot be collected in other ways. The problem with this solution is that, it requires an extra level of trust -- investigators must trust the management plane. Virtual Machine Introspection (VMI) can also be helpful in forensic investigation. Using this process, the investigators can execute a live forensic analysis of the system, while keeping the target system unchanged \cite{hay2008forensics}. 

Though the above solutions are important for cloud forensics, none of the existing research efforts in cloud forensics address the problem of proving past data possession. In this paper, we take the first step towards providing a solution for the past data possession. Combining all the previous solutions and our scheme will drive towards building a forensic-enabled cloud.

%% file: conclusion.tex
\section{Conclusion and Future Work}
\label{sec:conclusion}
With the increasing use of cloud computing, cloud forensics has attracted the attention of the security and forensics research community. Researchers have explored the challenges and proposed some solutions to mitigate the challenges. Several researchers proposed continuous synchronization of cloud data to overcome the forensic investigation challenges, though no scheme has been proposed yet about continuous synchronization. PPDP can be the solution of continuous synchronization. It will allow users and investigators to verify the possession of incriminating evidence by a suspect at a past time. In this paper, we proposed a scheme of PPDP, described how it can be utilized in cloud forensics, and provided the performance of our scheme by a proof-of-concept application. Our experimental result indicates that the overhead associated with generating the proof of evidence is very low and practically implementable by the CSP.

In future, we will implement our scheme on our own cloud platform using a open source framework. By implementing our private cloud platform, we can perform more sophisticated tests and will be able to make a practically usable application. After our successful implementation, we will collaborate with a CSP to deploy the scheme in a real-life cloud infrastructure.

%% file: ack.tex
\section*{Acknowledgments}
This research was supported by a Google Faculty Research Award, the Office of Naval Research Grant  \#N000141210217, the Department of Homeland Security Grant \#FA8750-12-2-0254, and by the National Science Foundation under Grant \#0937060 to the Computing Research Association for the CIFellows Project.